# Generic Pipelined Processor Modeling and High Performance Cycle-Accurate Simulator Generation

Mehrdad Reshadi, Nikil Dutt
Center for Embedded Computer Systems (CECS),
Donald Bren School of Information and Computer Science,
University of California Irvine, CA 92697, USA.
{reshadi, dutt}@cecs.uci.edu

## Abstract

*Detailed modeling of processors and high performance cycle-accurate simulators are essential for today's hardware and software design. These problems are challenging enough by themselves and have seen many previous research efforts. Addressing both simultaneously is even more challenging, with many existing approaches focusing on one over another. In this paper, we propose the Reduced Colored Petri Net (RCPN) model that has two advantages: first, it offers a very simple and intuitive way of modeling pipelined processors; second, it can generate high performance cycle-accurate simulators. RCPN benefits from all the useful features of Colored Petri Nets without suffering from their exponential growth in complexity. RCPN processor models are very intuitive since they are a mirror image of the processor pipeline block diagram. Furthermore, in our experiments on the generated cycle-accurate simulators for XScale and StrongArm processor models, we achieved an order of magnitude (~15 times) speedup over the popular SimpleScalar ARM simulator.*

## 1. Introduction

Efficient and intuitive modeling of processors and fast simulation are critical tasks in the development of both hardware and software during the design of new processors or processor based SoCs. While the increasing complexity of processors has improved their performance, it has had the opposite effect on the simulator speed. Instruction Set Simulators simulate only the functionality of a program and hence, enjoy simpler models and well established high performance simulation techniques such as compiled simulation and binary translation. On the other hand, cycle-accurate simulators simulate the functionality and provide performance metrics such as cycle counts, cache hit ratios and different resource utilization statistics. Existing techniques for improving the performance of cycle-accurate simulators are usually very complex and sometimes domain or architecture specific. Due to the complexity of these techniques and the complexity of the architecture, generating retargetable high performance cycle-accurate simulators has become a very difficult task.

To avoid redevelopment of new simulators for new or modified architectures, a retargetable framework uses an architecture model to automatically modify an existing simulator or generate a customized simulator for that architecture. Flexibility and complexity of the modeling approach as well as the simulation speed of generated simulators are important quality measures for a retargetable simulation framework. Simple models are usually limited and inflexible while generic and complex models are less productive and generate slow simulators. A reasonable tradeoff between complexity, flexibility and simulation speed of the modeling techniques has been seldom achieved in the past. Therefore, automatically generated cycle-accurate simulators were more limited or slower than their manually generated counterparts.

*Colored Petri Net* (CPN) [1] is a very powerful and flexible modeling technique and has been successfully used for describing parallelism, resource sharing and synchronization. It can naturally capture most of the behavioral elements of instruction flow in a processor. However, CPN models of realistic processors are very complex mostly due to incompatibility of a token-based mechanism for capturing data hazards. Such complexity reduces the productivity and results in very slow simulators. In this paper, we present *Reduced Colored Petri Net* (*RCPN*), a generic modeling approach for generating fast cycle-accurate simulators for pipelined processors. RCPN is based on CPN and reduces the modeling complexity by redefining some of CPN concepts and also using an alternative approach for describing data hazards. Therefore, it is as flexible as CPN but far less complex and can support a wide range of architectures. Figure 1 illustrates the advantages of our approach using an example pipeline block diagram and its corresponding RCPN and CPN models. It is possible to convert an RCPN to a CPN and hence reuse the rich varieties of analysis, verification and synthesis techniques that have been proposed for CPN. The RCPN is intuitive and closely mirrors the processor pipeline structure. RCPN provides necessary information for generating fast and efficient cycle-accurate simulators. For instance, our XScale [3] processor cycle-accurate simulator runs an order of magnitude (~15 times) faster than the popular SimpleScalar simulator for ARM [2].

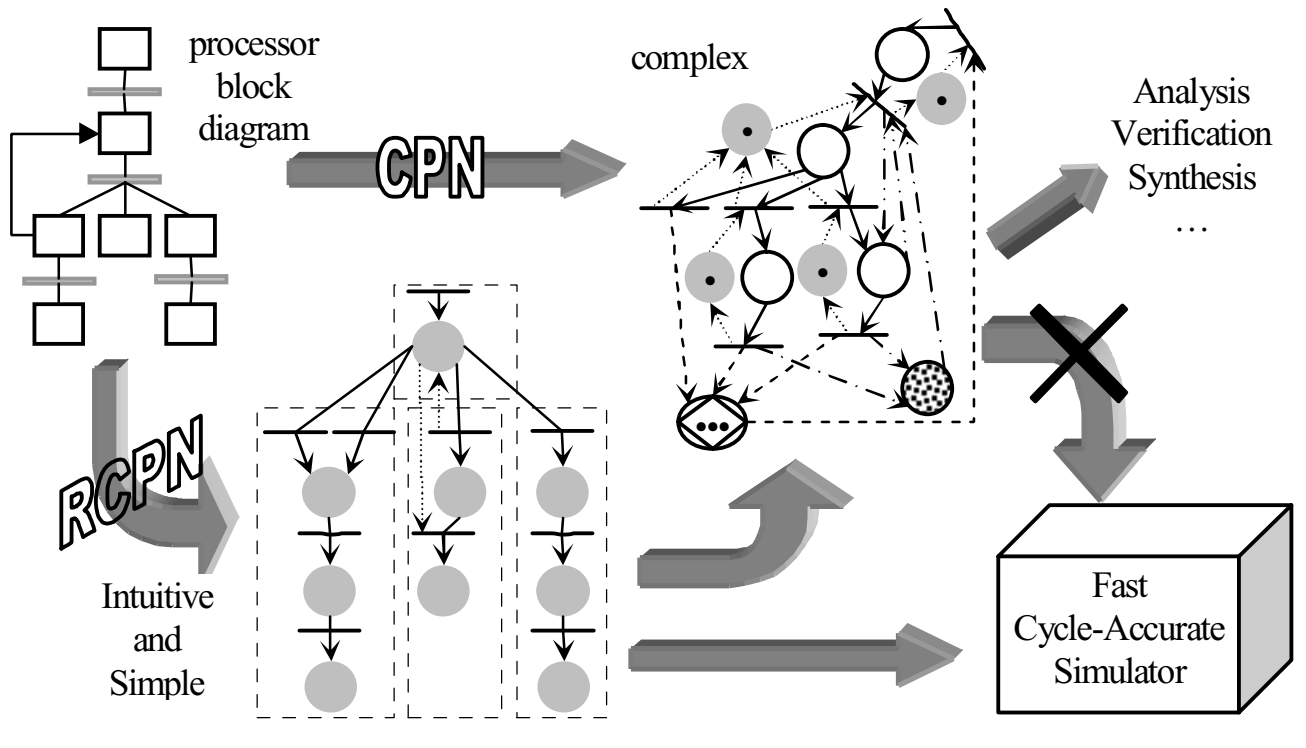

**Figure 1- Advantages of RCPN: Intuitive, Fast Simulation**

In this paper, Section 2 summarizes the related works. Section 3 describes the RCPN model and illustrates the details of the pipeline example of Figure 1. Section 4 explains the simulation engine and optimizations that are possible because of RCPN. Section 5 shows the experimental results and Section 6 concludes the paper.

## 2. Related Work

Detailed micro-architectural simulation has been the subject of active research for many years and several models and techniques have been proposed to automate the process and improve the performance of the simulators.

ADL based approaches such as ISDL [7], nML [6], and EXPRESSION [8] take an operation-centric approach and automate the generation of code generators. These ADLs describe instruction behaviors in terms of basic operations, but do not explicitly support detailed pipeline control-path specification which limits their flexibility for generating micro-architecture simulators. The Sim-nML [9] language is an extension to nML to enable cycle-accurate modeling of pipelined processors. It generates slow simulators and cannot describe processors with complex pipeline control mechanisms due to the simplicity of the underlying instruction sequencer.

Hardware centric approaches, such as BUILDABONG [11] and MIMOLA [10], model the architectures at the register transfer level and lower levels of abstraction. This level of abstraction is not suitable for complex microprocessor modeling and results in very slow cycle-accurate simulators. Similarly, ASim [12] and Liberty [13] model the architectures by connecting hardware modules through their interfaces. Emphasizing reuse, they use explicit port-based communication which increases the complexity of these models and have a negative effect on the simulation speed.

SimpleScalar [2] is a tool-set with significant usage in the computer architecture research community and its cycle-accurate simulators have good performance. It uses a fixed architectural model with limited flexibility through parameterization. Babel [14] was originally designed for retargeting the binary tools and has been recently used for retargeting the SimpleScalar simulator. Running as fast as SimpleScalar, UPFAST [15] takes a hardware centric approach and requires explicit resolution of all pipeline hazards. Chang et al [16] have proposed a hardware centric approach that implicitly resolves pipeline hazards in the cost of an order of magnitude slow down in simulation performance. FastSim [17] uses the Fast-Forwarding technique to perform an order of magnitude (5-12 times) faster than SimpleScalar. Fast-Forwarding is one of very few techniques with such a high performance; however, generating simulators based on this technique is very complex. To decrease this complexity, Facile [18] has been proposed to automate the process. But the automatically generated simulators suffer significant loss of performance compared to FastSim and run only 1.5 times faster than SimpleScalar. Besides, modeling in Facile requires more understanding of the Fast-Forwarding technique rather than the actual hardware being modeled.

LISA [19] uses the L-chart formalism to model the operation flow in the pipeline and simplifies the handling of structural hazards. It has been used to generate fast retargetable compiled simulators. The flexibility of LISA is limited by the L-chart formalism and handling behaviors, such as data hazards, requires explicit coding in C language. Operation State Machine (OSM) [20] models a processor in two layers: hardware layer and operation layer. The hardware layer captures the functionality and connectivity of hardware components, simulated by a discrete event simulator. The operation layer captures the flow of operations in the hardware using Finite State Machines (FSMs). The FSMs communicate with the underlying hardware components by exchanging tokens (events) through Token Manager Interfaces (TMI), which define the meaning of these tokens. The generated simulators in this model run as fast as SimpleScalar.

Petri Nets have also been used for modeling processors [22]. They are very flexible and powerful and additionally, allow several formal analyses to be performed. Simple Petri Net models are also easy to visualize. Colored Petri Nets [1] simplify the Petri Nets by allowing the tokens to carry data values. However, for complex designs, as that of processor pipeline, their complexity grows exponentially which makes modeling very difficult and significantly reduces simulation performance. In this paper, we propose the Reduced Colored Petri Net (RCPN) model which benefit from Petri Net features while being simple and capable of deriving high performance cycle-accurate simulators. It is an instruction-centric approach and captures the behavior of instructions in each pipeline stage at every clock cycle via modified and enhanced Colored Petri Net concepts.

## 3. Reduced Coloured Petri Net

To describe the behavior of a pipelined processor, operation latencies and data, control and structural hazards must be captured properly. A token based mechanism, such as CPN, can easily model variable operation latencies and basic structural and control hazards. Because of architectural features such as register overlapping, register renaming and feedback paths, capturing data hazards using a token based mechanism is very complex and difficult. In RCPN, we redefine the concepts of CPN to make it more suitable for processor modeling and fast simulation. As for the data hazards, we use a separate mechanism that is explained in Section 3.1.

Figure 2(a) shows a very simple pipeline structure with two latches and four units and Figure 2(b) shows the CPN model that captures its structural hazards. In this figure, circles show *places* (states), boxes show *transitions* (functions) and black dots represent *tokens*. In CPN, a transition is enabled when it has one token of proper type on each of its input arcs. An enabled transition can *fire* and remove tokens from its input places and generate tokens for its output places. In this pipeline, if latch $L_2$ is available and a proper instruction is in latch $L_1$, then the functionality of unit $U_2$ is executed, $L_2$ is occupied and $L_1$ becomes available for next instruction. This behavior is represented by the availability of tokens in Figure 2(b). Here, whenever $U_1$ is enabled, it removes the token of $L_1$ and puts it in $P_1$. Then, if $L_2$ has a token, $U_2$ can fire and move a token from $P_1$ to $L_1$ and from $L_2$ to $P_2$. In other words, whenever a token is in place $L_1$, it means that latch $L_1$ in Figure 2(a) is available and unit $U_1$ can send a new instruction into it; and whenever a token is in place $P_1$, it means that an instruction is available in latch $L_1$ and unit $U_2$ or $U_4$ can use it. The back-edges (dotted lines) create circular loops. The number and complexity of these loops in the CPN of a typical pipeline grows very rapidly with its size. These loops not only make the CPN models very complex, they are also the main obstacle in generating high performance simulators.

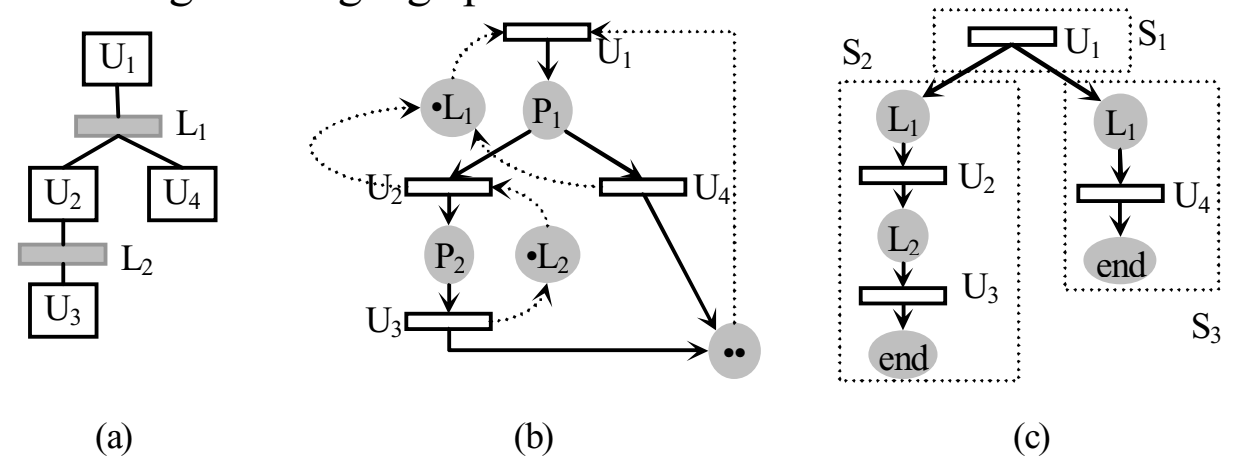

**Figure 2- An Example Pipeline structure(a) and its CPN(b) and RCPN(c) models**

The RCPN model is based on the same concept as CPN; i.e. when a *transition* is enabled, it fires and removes *tokens* from the input *places* and generates tokens for the output places. In RCPN processor models, structural hazards, control hazards and variable operation latencies are naturally modeled by tokens, places, transitions and delays. To simplify processor models, we redefine these concepts in RCPN as follows:

**Places:** A place shows the state of an instruction. To each place a pipeline stage is assigned. A pipeline stage is a latch, reservation station or any other storage element in the pipeline that an instruction can reside in. For each pipeline stage that an instruction may go through, there will be at least one place in the model. Each pipeline stage has a *capacity* parameter that determines how many tokens (instructions) can reside in it at any time. We assume when instructions finish they go to a final virtual pipeline stage, called *end*, with unlimited capacity. The places to which this virtual final stage is assigned represent the final state of the corresponding instructions. In RCPN, each place is shown with a circle in which the name of the corresponding pipeline stage is written. Places with similar name share the capacity of their pipeline stage. The tokens of a place are stored in its pipeline stage.

**Transition:** A transition represents the functionality that must be executed when the instruction changes its state (place). This functionality is executed (fired) when the transition is enabled. A transition is enabled if its *guard* condition is true and there are enough tokens of proper types on its input arcs AND the pipeline stages of the output places have enough capacity to accept new tokens. A transition can directly reference non-pipeline units such as branch predictor, memory, cache etc. The transition may use the functionality of these units to determine the type, value and delay of tokens that it sends to its output places.

**Arc:** An arc is a directed connection between a place and a transition. An arc may have an expression that converts the set of tokens that pass through the arc. For deterministic execution, each output arc of a place has a priority that shows the order at which the corresponding transitions can consume the tokens and become enabled.

**Token:** There are two groups of tokens: *reservation* tokens that carry no data and their presence in a place indicates the occupancy of the place's corresponding pipeline stage; and *instruction* tokens that carry complex data depending on the type of the instruction.

Instruction tokens are the main focus of the model since each instruction token represents an instruction being executed in the pipeline. In other words, RCPN describes how an individual instruction flows through stages of the pipeline. In any RCPN, there is one instruction independent sub-net that generates the instruction tokens, and for each instruction type, there is a corresponding sub-net that distinctively describes the behavior of instruction tokens of that type. Figure 2(c) shows the RCPN model of the simple pipeline shown in Figure 2(a). The model is divided into three sub-nets: $S_1$, $S_2$ and $S_3$. $S_1$ describes the instruction independent portion that generates two types of instruction tokens. Note that as long as state $L_1$ has room for a new token, transition $U_1$ can fire. In fact, because of our new definition of "transition enable", an RCPN model can start with a transition as well as a place. Any sub-net can generate an instruction token and send it to its corresponding sub-net. This is equivalent with instructions that generate multiple micro operations in a pipeline (e.g. Multiple LoadStore instruction in XScale). As in real processor, instruction tokens never go through circular paths[1].

A delay may be assigned to a place, a transition or a token. These delays have default values and can be changed at run time based on data values, etc. The delay of a place determines how long a token should reside in that place before it can be considered for enabling an output transition. The delay of a transition expresses the execution delay of the functionality of that transition. The delay of a token overwrites the delay of its containing place and has the same effect. By changing the delay of a token, a transition can indirectly change the delay of its output place.

Usually in microprocessors, the instructions that flow through a similar pipeline path have similar binary format as well. In other words, the instructions that go through the same functional units have similar fields in their binary format. Therefore, a single decoding scheme and behavior description can be used for such group of instructions which we refer to as an *Operation Class*. An operation class describes the corresponding instructions by using symbols to refer to different fields in their binary code. A symbol can refer to a Constant, a μ-operation or a Register. Using these symbols, for each operation class an RCPN sub-net describes the behavior of the corresponding instructions. During instruction decode, the actual values of these symbols are determined. Therefore, by replacing the symbols with their values, a customized version of the corresponding RCPN sub-net is generated for individual instances of instructions. Figure 4(b) shows examples of such operation classes. The details of using symbols and the decode algorithm are described in [4].

## 3.1 Capturing Data Hazards

To capture data hazards, we need to know when registers can be read or updated and if an instruction is going to update a register, what its state is at any time. In many processors, registers may overlap[2] and hence modifying one may affect the others. On the other hand, generally instructions use different pipeline stages to read source operands, calculate results, or update destination operands. Therefore, instructions must be able to hold register values after reading or before updating registers.

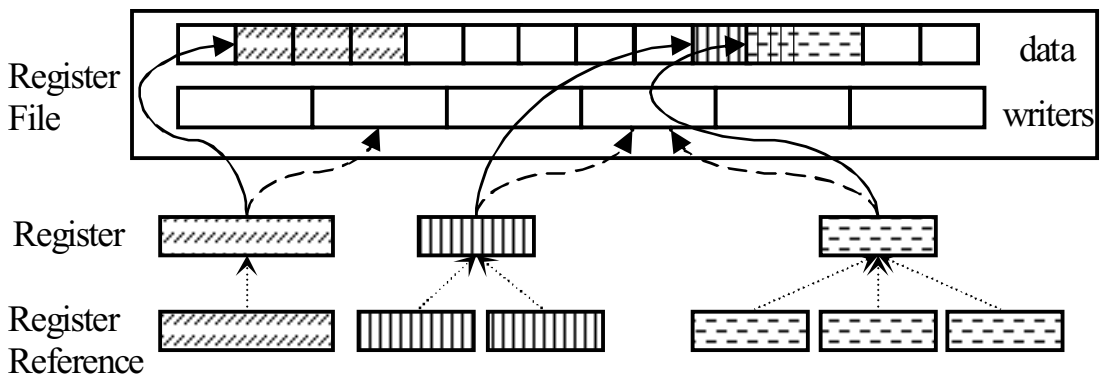

**Figure 3- Register structure**

Addressing all the above issues with a token based mechanism is very complicated and hence, in RCPN we use an alternative approach that explicitly supports a lock/unlock (semaphore) mechanism for accessing registers, temporary locations for register values, and registers with overlapping storage for data. As Figure 3 shows, we model registers at three levels:

**Register File:** It defines the actual storages for data, register renaming and pointer to instructions that will write to a register. There may be multiple register files in a design.

**Register:** Each register has an index and points to proper storages of the register file. Multiple registers, can point to the same storage areas to represent overlapping.

**Register Reference (RegRef):** Each RegRef points to a register and has an internal storage for storing the register value. A symbol in an operation class that points to a register is replaced by a proper RegRef during decode. In fact, RegRefs represent the pipeline latches that carry instruction data in real hardware. During simulation, this is almost equivalent with renaming registers for each individual instruction. RegRefs' internal values are used in the computations and the instructions access and update registers through RegRefs' interfaces. The interface is fixed and includes: *canRead*(), true if register is ready for reading; *canRead*(*s*), true if the instruction that is going to update the corresponding register is in state *s* at the time of call; *read*(), reads the values of corresponding register and stores it in the internal storage of RegRef; *canWrite*(), true if the register can be written; *reserveWrite*(), assigns the current RegRef pointer and its containing instruction as the writers of the corresponding register; *writeback*(), writes the internal value of the RegRef to the corresponding register and may reset its writer pointers; and *read*(*s*), instead of reading the value of the corresponding register, it reads the internal value of the writer RegRef whose containing instruction is in state *s* at the time of call. The *read*(*s*) interface provides a simple and generic means of modeling data forwarding through feedback or bypass paths.

[1] A token may stay in one stage and produce multiple tokens to go through the same path and repeat a set of behaviors.

[2] E.g. overlapping register-banks in ARM or register windows in SPARC.

In RCPN, data hazards are explicitly captured by using Boolean interfaces, such as *canRead*, in the arcs' guard conditions; and using normal interfaces, such as *read*, in the transitions. These pairs of interfaces must be used properly to ensure correctness of the model. Whenever *read*(), *reserveWrite*() or *read*(*s*) appears in a transition, *canRead*(), *canWrite*() or *canRead*(*s*) must appear in the guard condition of its input arc, respectively.

The implementation of these interfaces may vary based on architectural features such as register renaming. For example, in a typical implementation of these interfaces, transition $T_1$ first checks *r.canWrite()* to check write-after-write and write-after-read hazards for accessing register *r*. Then it calls *r.reserveWrite()* to prevent future reads or writes. After calling *r.writeback()*in another transition, register *r* can be safely read or written. In RCPN, a symbol in an operation class that points to a constant is replaced by a *Const* object during decode. The Const object provides the same interface as of RegRef with proper implementation. For example, its *canRead*() always returns true; its *writeback*() does nothing and so on. In this way, data hazards can be uniformly captured using symbols in the operation class.

The next section demonstrates most of the RCPN modeling capabilities via an example.

## 3.2 Example RCPN Processor Model

Figure 4(a) shows the block diagram of a representative out-of-order completion processor with a feedback path. Figure 4(b) shows three types of instructions (operation classes) in this processor. Each instruction consists of symbols whose actual value is determined during instruction decode. For example, the *L* symbol in *LoadStore* is a Boolean symbol and is true for loads and false for store instructions. To show the flexibility of the model, we assume that the feedback path is used only for the first source operand of *ALU* instructions ($s_1$).

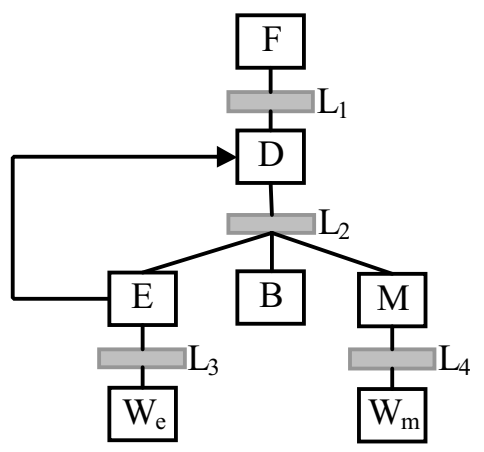

```
Branch {
   offset: {Register | Constant}
};
ALU {
   op: {Add | Sub | Mul | Div | …}
   d, s1: {Register}
   s2 : {Register | Constant}
};
LoadStore {
   L: {true | false}
   r: {Register}
   addr: {Register | Constant}
};
```

(a) (b)

**Figure 4- Representative out-of-order processor**

Figure 5 shows the complete RCPN model of the above processor. It contains one instruction independent sub-net and three instruction specific sub-nets. The boxes show the functionality of transitions and the codes above them show their guard conditions. The guard conditions are written in the form of [$cond_1$, $cond_2$ …] which is equivalent with: $cond_1 \wedge cond_2 \wedge \ldots$

To model the feedback path, two arcs with different priorities come out of place $L_1$ and enter the ALU instruction sub-net. If the first arc, with priority 0, cannot read the value of first source operand, then the second arc, with priority 1, verifies that the writer instruction of operand $s_1$ is in the state $L_3$ and then reads it. Otherwise, the instruction is stalled in $L_1$. After reading the source operand and reserving the destination for writing, the result is calculated in transition *E* and stored in the internal value of the destination *d*. This value is finally written back in transition $W_e$.

In Branch instruction sub-net, the dotted arcs represent reservation tokens. Therefore in this example, when a branch instruction is issued, it stalls the fetch unit by occupying latch $L_1$ with a reservation token and disabling the fetch transition. In the next cycle, this token is consumed and the fetch unit is un-stalled. An alternative implementation is flushing $L_1$ and $L_2$ latches in transition *B* instead of using reservation tokens.

The LoadStore instruction sub-net demonstrates the use of token delay in transition *M* to model the variable delay of memory (cache). It also shows how data dependent delays can be modeled in RCPN. The component *mem*, referenced in this transition, can be used from a library or reused from other designs.

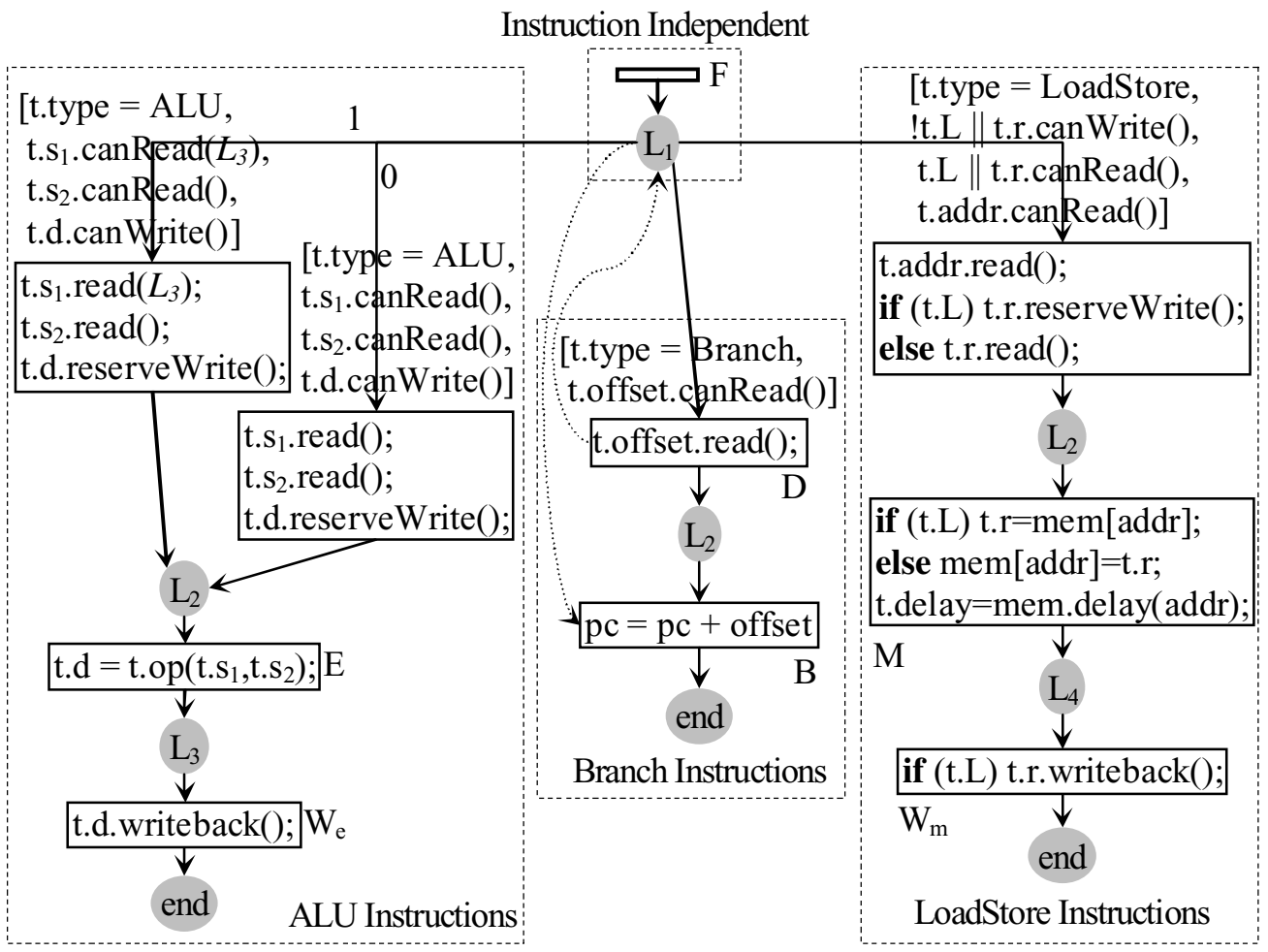

**Figure 5- RCPN sub-nets**

Processor RCPN models can be converted to standard CPN and use all the tools and algorithms that is available for CPN. Details of this conversion and more complex examples capturing VLIW and multi-issue machines as well as RCPN model of the Tomasulo algorithm are detailed in our technical report [5].

# 4. Cycle-accurate Simulation

RCPN can generate very fast cycle-accurate simulators. Like any other Petri Net model, an RCPN model can be simulated by locating the enabled transitions and executing them concurrently. Searching for enabled transitions and handling concurrency can be very time consuming in generic Petri Net models especially if there are too many places and transitions in the design. However, a more careful look at the RCPN model reveals some of its properties that can be utilized to simplify these two tasks and speed up the simulation significantly.

Of the two groups of tokens in RCPN, reservation tokens carry no data and are used only to show unavailability of resources. Since transitions represent the functionality of an instruction between two pipeline stages, reservation tokens alone can not enable them. Therefore, only places that have an instruction token may have an enabled output transition. While a place may be connected to many transitions in different sub-nets, an instruction token only goes through transitions of the sub-net corresponding to its type. In other words, based on the type of an instruction token, only a subset of output transitions of a place may be enabled. Since the structure of RCPN model is fixed during simulation, for every place and instruction type the list of transitions that may be enabled can be statically extracted from the model before simulation begins. This list is sorted based on the priorities of output arcs of the place and processed accordingly. Figure 6 shows the pseudo code that extracts this list for each place in RCPN and each instruction type in the ISA and stores it in *sorted_transitions* table. This code is called before program simulation begins and hence has no runtime overhead for simulation.

```
CalculateSortedTransitions(){
Arcs = {(p, t), (t, p)| p∈ Places and t ∈ Transitions};
foreach place p in P
  foreach InstrutionType IType in Instruction-Set
    sorted_transitions[p, IType]=(t0, t1, ...) such that
      (p, ti) ∈ Arcs, ti ∈ subnet(IType),
      i < j => priority( (p, ti) ) < priority( (p, tj) );
  endfor
endfor
}
```

**Figure 6-Extracting and sorting transition subsets**

Figure 7 shows the pseudo code for processing the output transitions of a place. It is called in each clock cycle to process the instructions that are in a particular state (place $p$). For each instruction, it finds the first transition that can be executed and move the instruction to its next state. The corresponding transitions list is looked up from the *sorted_transitions* table.

```
Process(place p){
  foreach instruction token inst in p
    foreach transition t in sorted_transitions[p, inst.type]
      if enabled(t)
        remove tokens from input places of t;
        execute transition function of t;
        add tokens to output places of t;
        break; //process next instruction token
      endif
    endfor
  endfor
}
```

**Figure 7-Processing places with instruction tokens**

In RCPN, enabled transitions execute in parallel; tokens are simultaneously read from input places at the beginning of a processor cycle, and then, in parallel, written to the output places at the end of the cycle. Therefore, the simulator must ensure that the variables representing such places are all read before being written during a cycle. The usual, and computationally expensive solution, is to model such places using a two-list algorithm (similar to master/slave latches). This approach uses two token storages per place- one of them is read from, and the other written to in the course of a cycle. At the end of the cycle, the tokens in the written-to storage are copied to the read-from storage.

In general, we can ensure that all tokens from the previous cycle are read-from before being written-to by evaluating all places (or their corresponding pipeline stages) in reverse topological order. Therefore, only very few places that are referenced in a circular way, usually because of feedback paths like state $L_3$ in Figure 5, need to implement a two-list algorithm. The resulting code is considerably faster since it avoids the overheads of managing two storages in the two-list algorithm. Note that in CPN, this well-known optimization is not applicable because all resource sharings are modeled with circular loops of places.

```
CalculatingSortedTransitions();
P = list of places in reverse topological order;
while program not finished
  foreach place p in {places that implement two-list algorithm}
    mark written tokens as available for read in p;
  endfor
  foreach place p in P
    Process(p);
  endfor
  execute the instruction independent sub-net of RCPN;
  increment cycle count;
endwhile
```

**Figure 8-Main body of simulation engine**

Figure 8 shows the main body of our simulation engine. In the main loop, after updating the places that implement the two-list algorithm, all places are processed in reverse topological order. At the end of each iteration, the instruction independent sub-net of the model, which is responsible for generating the instruction tokens, is executed.

## 5. Experiments

To evaluate the RCPN model, we modeled both StrongArm [21] and XScale [3] processors using the ARM7 instruction set. StrongArm has a simple five stage pipeline. XScale is an in-order execution, out-of-order completion processor with a relatively complex pipeline structure shown in Figure 9. The ARM instruction set was implemented using six operation-classes [4]. Using these operation classes, it took only one man-day for StrongArm and only three man-days for XScale to develop both the RCPN models and the simulators.

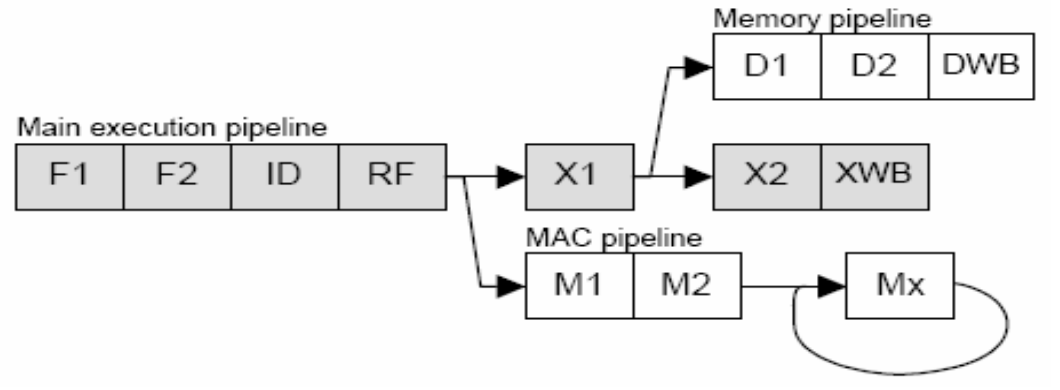

**Figure 9-XScale pipeline**

To evaluate the performance of the simulators we chose benchmarks from MiBench [23] (blowfish, crc), MediaBench [24] (adpcm, g721) and SPEC95 [25] (compress, go) suites. These benchmarks were selected because they use very few simple system calls (mainly for IO) that should be translated into host operating system calls in the simulator. We used *arm-linux-gcc* to generate the binary code of the benchmarks. The compiler only uses ARM7 instruction-set and therefore we only needed to model those instructions. The simulators were run on a Pentium 4/1.8 GHz/512 MB RAM.

Figure 10 compares the performance of the simulators generated from RCPN model with that of SimpleScalarArm. The first bar for each benchmark shows the performance of SimpleScalarArm simulator. This simulator implements StrongArm architecture and we disabled all checkings and used simplest parameter values to improve simulation performance. On the average this simulator executes 600k cycles/sec. The second and third bar for each benchmark shows the performance of our simulator for XScale and StrongArm processor models respectively. These simulators execute 8.2M cycles/sec and 12.2M cycles/sec on the average. SimpleScalar uses a fixed architecture for any processor model. Therefore, the complexity and performance of the simulator is similar across different models. On the other hand, RCPN models are true to the modeled processor and hence the complexity of generated simulators depends on the complexity of the processor that they simulate. Due to its simpler pipeline, the StrongArm simulator performs better that that of XScale.

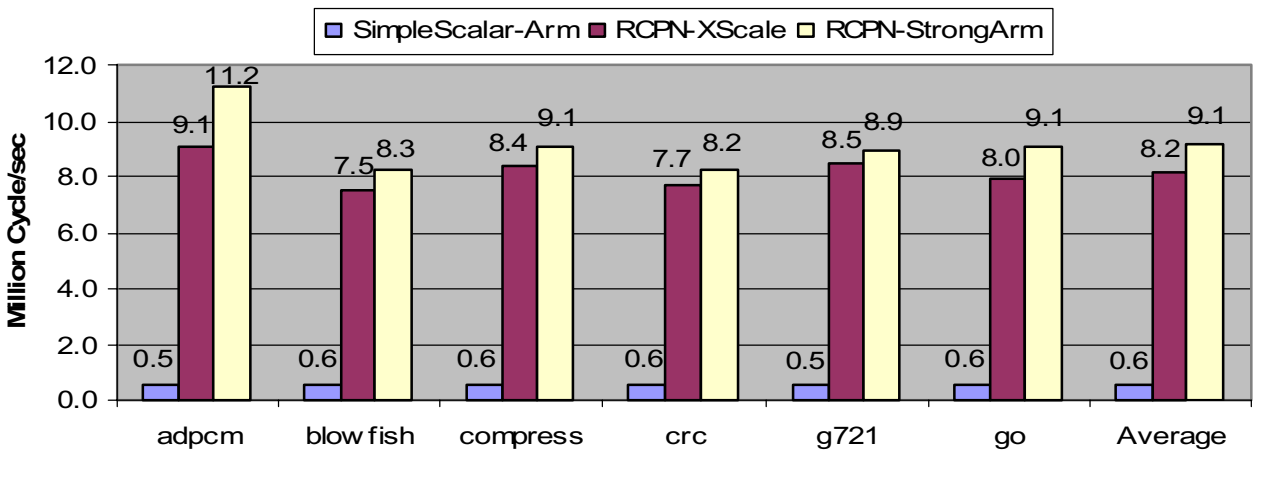

**Figure 10-Simulation performance (Million cycle/second)**

Figure 11 compares the CPI values of SimpleScalarArm and our StrongArm simulator. This figure shows that although our simulator runs significantly faster than SimpleScalar, the CPI values of the two simulators are almost similar. The ~10% difference is due to the accuracy of the information in the model used for generating the simulator. The RCPN based modeling approach does not impose any limitation on capturing instruction schedules. Therefore, by providing accurate models, the results of generated simulators can be fairly accurate. Such models are usually obtained by comparing the simulation results against either a base simulator or the actual hardware, and then refining the model information.

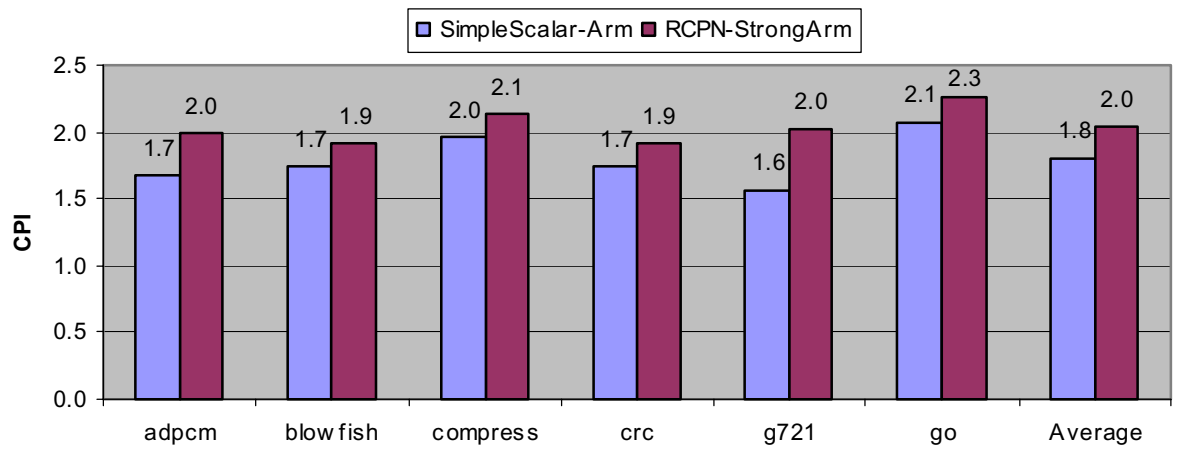

**Figure 11-Clocks per instruction (CPI)**

From modeling capability point of view, RCPN and OSM are comparable. However, OSM uses very few FSMs, e.g. only one FSM for StrongARM, and captures the pipeline through these FSMs and TMI software components. RCPN uses multiple sub-nets, each equivalent with an OSM, to explicitly capture the pipeline control. For example, there are six RCPN sub-nets in the StrongArm model. Only for capturing data hazards, RCPN relies on the fixed interface software components. Therefore, a larger part of processor behavior is captured formally in RCPN than in OSM. In other words, the non-formal part of OSM model (TMIs) is large enough that it needs a separate event-driven simulation engine; but the non-formal part of RCPN model is a set of very simple functions for accessing registers. Nevertheless, RCPN based simulators run an order of magnitude faster than OSM based ones. Our simulators are as fast as FastSim while we use two simple optimizations and FastSim uses the very complex Fast-Forwarding technique. We can summarize the reasons of this high performance as follows:

- Because of RCPN features, we can reduce the overheads of supporting concurrency and searching for enabled transitions.
- We apply partial evaluation optimization to customize the instruction dependent sub-nets for each instruction instance and hence improve their performance.
- In RCPN, when an instruction token is generated, the corresponding instruction is decoded and stored in the token. Since the token carries this information, we do not need to re-decode the instruction in different pipeline stages to access its data. Furthermore, the tokens are cached for later reuse in the simulator.

Since the simulator is generated automatically, debugging of the implementation of the simulator is (eventually) eliminated. Only the model itself must be debugged/verified. As [4] describes, using operation-classes and templates make debugging much simpler. Since RCPN is formal and can be converted to standard CPN, formal methods also can be used for analyzing the models.

# 6. Conclusion

In this paper, we presented the RCPN model for capturing pipelined processors. RCPN benefits from the same concepts as other Petri Net models and has two advantages: first, it provides an efficient way for modeling architectures; and second, it generates high performance cycle accurate simulators. RCPN models are very intuitive to generate because they are very similar to the pipeline block diagram of the processor. Our cycle-accurate simulators, for both StrongArm and XScale processors, run about 15 times on average faster than SimpleScalar for ARM, although XScale has a relatively complex pipeline.

The use of Colored Petri Net concepts in RCPN makes it very suitable for different design analysis and verification purposes. The clean distinction between different types of tokens and data hazard mechanism in addition to the structure of RCPN can be used to extract the necessary information for deriving retargetable compilers. The future direction of our research is to address these issues as well as extracting fast functional simulators from the same detailed RCPN models.

# 7. Acknowledgement

This work was partially supported by NSF grants: CCR-0203813 and CCR-0205712.